\documentstyle{article}
\input psfig
\begin{document}
\title{Nucleon-Antinucleon Interaction from the Skyrme Model: II. \\
Beyond the Product Ansatz}

\author{Yang Lu, Pavlos Protopapas and R.~D.~Amado \\
Department of Physics and Astronomy\\
University of Pennsylvania,
Philadelphia, Pennsylvania 19104}

\date{September 13, 1997}

\maketitle
\begin{abstract}
We calculate the full static interaction of a Skyrmion
and an anti-Skyrmion as a function of separation and relative grooming.
From this, using projection methods and Born-Oppenheimer mixing,
we obtain the nucleon antinucleon interaction. We find agreement with 
the major
features of the empirical interaction including the strong central 
attraction and the sharp onset of annihilation at about 1 fm. 
\end{abstract}

\section{Introduction}

Making a theory of low energy nucleon-antinucleon annihilation from the
entrance channel through to the annihilation products presents a daunting
challenge. An exact QCD based calculation of the  
process is beyond our 
present analytical understanding and our computational resources. 
We are left then with either abandoning the problem or using some
approximate or effective theory. Such a theory must include an account 
of the nature of the nucleon and the antinucleon, in order to describe
annihilation, as well as containing a description of their interaction. 
The  theory should also have its roots in QCD. The only such
effective theory of annihilation that contains 
the requisite ingredients and gives hope
of being tractable is based on the Skyrme model \cite{Skyrme}.
This models QCD in the classical or large number of colors
($N_C$) limit and at long distances or low momentum \cite{'tHooft,Witten}.  
Sommermann et al. \cite{SSLK} have  
shown, in a detailed numerical investigation, that a Skyrme treatment of
annihilation gives great insight into the process, and we have exploited
that insight to show that the Skyrme treatment can account for the 
major features of the annihilation channels \cite{batch2}.

Thus far most of our results have come from a very simple picture of the
the annihilation final state. We start with a spherical ``blob" of 
Skyrmionic matter with zero baryon number and energy of two nucleon masses.
We propagate this blob using the classical Skyrme equations, 
including classical vector meson fields, until it is
well into the radiation zone and then use coherent states to reintroduce
the meson quanta of the final state. 
 This gives a remarkably 
good overall picture of the observed branching ratios seen in low energy
nucleon antinucleon annihilation. However it does not give any insight into
the incoming nucleon antinucleon channel. For this we have to study the
initial state dynamics. A full dynamical Skyrmion calculation of the 
initial state is difficult and fraught with instabilities \cite{SSLK,Livermore}.
 Therefore we take a simpler approach of using the 
Skyrme dynamics to obtain the nucleon antinucleon interaction potential.
From that we later can connect the entrance channel to the annihilation blob
and from there proceed as before. 

Our previous attempt to obtain the nucleon antinucleon interaction from the 
Skyrme model was based on the product ansatz \cite{LuRDA96}. This calculation
reproduces some of the major features of the observed interaction, but
fails to give the observed strong central attraction. Such a failure is
also present in the product ansatz approach to the nucleon nucleon interaction,
\cite{VinhMau}, and is a failing of the ansatz, not of the Skyrme approach.
In the nucleon nucleon case the full central attraction was obtained by taking
two steps. First a full dynamical calculation of the interaction energy
of two Skyrmions as a function of separation and relative grooming was 
done \cite{W&W}. This is a difficult and numerically intensive calculation.
Then the results of that calculation were combined with calculation of
dynamical $\Delta$ mixing using the Born-Oppenheimer method, \cite{WA},
to yield a nucleon nucleon interaction with most of the empirical features,
in particular the strong central attraction. In this paper we report 
a calculation of the nucleon antinucleon interaction taking the corresponding
two steps. First we compute the Skyrmion anti-Skyrmion interaction energy 
as a function of separation and relative grooming using the full Skyrme
dynamics. As in the Skyrmion Skyrmion case, this is a difficult 
and complex computation. We then use Born-Oppenheimer techniques as well
as standard Skyrmion to baryon projection methods to
obtain the nucleon antinucleon potentials. We find strong central 
attraction between the nucleon and antinucleon of the form and magnitude
seen from the data, 
as well as qualitative agreement with the other principal features 
of the empirical potentials. We also find strong and sudden  coupling of
the nucleons to the annihilation channel at about 1 fm, just as is seen
in empirical fits to the data. 

This paper is organized as follows. In the next section we present the major
features of the formulation we use for the dynamical calculation. We do not
go into great detail here since much of the formulation is already well 
presented in the literature.  The following section reports the numerical 
methods we used. Since the calculation is challenging and difficult, we
give some detail here. However the reader interested only in the finished
answers can skip this section. We follow with a section on results that
reports both our Skyrmion anti-Skyrmion interaction and our nucleon antinucleon
interaction as well as showing some plots of Skyrme fields. We end with a
discussion section that reviews briefly what our next steps might be. 

\section{Formulation}

The Skyrme model, \cite{Skyrme}, is a classical, nonlinear field theory 
of an SU(2) valued field. The Lagrangian for the theory is written
\begin{equation}
{\cal L}=-\frac{f_\pi^2}{4}{\rm Tr} \left[L_\mu L^\mu\right]
+\frac{1}{32e^2}{\rm Tr}\left[L_\mu,\;L_\nu\right]^2+\frac{m_\pi^2 f_\pi^2}{2}
{\rm Tr} \left[U-1\right],
\end{equation}
where $L_{\mu}$ is expressed in terms of the unitary SU(2) valued
matrix $U$ by
\begin{equation}
L_\mu=U^+\partial_\mu U .
\end{equation}
The first term of the Lagrangian comes from the nonlinear sigma
model. The second, the Skyrme term, was introduced by Skyrme to 
stabilize the model, and the third term is the pion mass term. It is
the term that breaks chiral invariance. The three constants in the 
theory are the pion decay constant, $f_{\pi}$, the constant $e$
(not the electric charge) introduced by Skyrme to set the scale of
his stabilizing term, and the pion mass, $m_{\pi}$. The Lagrangian 
admits topologically stable objects carrying a conserved winding
number that Skyrme associated with baryon number. It is customary
to adjust $f_{\pi}$ and $e$ to yield the correct long distance tail 
of the pion wave in the nucleon. 
We follow \cite{W&W} in this and take $f_{\pi} = 93$ MeV and 
$e=4.76$. We take the pion mass at its observed value\footnote{
Note how  few parameters are in the Skyrme model and note further that
we adjust none of them to calculate the interaction energy.}.  
The unitary SU(2) valued field $U$ can be written in terms of the
chiral angle iso-vector field $\mbox{\boldmath $F$}$ as follows,
\begin{equation}
U=\exp i\mbox{\boldmath $\tau$}\cdot \mbox{\boldmath $F$},
\end{equation}
and then $L_{\mu}$ expressed in terms of $\bf{F}$ by
\begin{equation}
L_\mu=i\mbox{\boldmath $\tau$} \cdot\mbox{\boldmath $A$ }_\mu,
\end{equation}
with 
\begin{equation}\label{Eq:A}
\mbox{\boldmath $A$}_\mu=\partial_\mu \mbox{\boldmath $F$}
\frac{\sin F \cos F}{F}
+\mbox{\boldmath $F$}\partial_\mu F \frac{F-\sin F\cos F }{F^2}+
\mbox{\boldmath $F$}\times \partial_\mu \mbox{\boldmath $F$}
\frac{\sin^2 F}{F^2}.
\end{equation}
With these definitions, the terms in the Skyrme Lagrangian become, 
for the nonlinear sigma term
\begin{eqnarray}
{\cal L}_\sigma &=& \frac{f_\pi^2}{2}\mbox{\boldmath $A$}_\mu
           \mbox{\boldmath $A$}^\mu \nonumber \\
        &=& \frac{f_\pi^2}{2}\left[\left(\partial_\mu 
 \mbox{\boldmath $F$}\cdot
\partial^\mu\mbox{\boldmath $F$}\right)\frac{\sin^2 F}{F^2}+
\left(\mbox{\boldmath $F$}  \cdot \partial_\mu\mbox{\boldmath $F$}\right)
\left(\mbox{\boldmath $F$}  \cdot \partial^\mu\mbox{\boldmath $F$}\right)
\frac{F^2-\sin^2 F}{F^4}
   \right] ,
\end{eqnarray}
where we used
\begin{equation}
\left[L_\mu, L_\nu\right]=-2i\mbox{\boldmath $\tau$}\cdot
   (\mbox{\boldmath $A$}_\mu\times \mbox{\boldmath $A$}_\nu),
\end{equation}
and 
\begin{equation}
{\rm Tr}\left[L_\mu, L_\nu\right]^2=-8\left(
\mbox{\boldmath $A$}_\mu\cdot \mbox{\boldmath $A$}^\mu 
\mbox{\boldmath $A$}
_\nu\cdot \mbox{\boldmath $A$}^\nu-\mbox{\boldmath $A$}_\mu
\cdot \mbox{\boldmath $A$}_\nu \mbox{\boldmath $A$}^\mu\cdot 
\mbox{\boldmath $A$}^\nu\right),
\end{equation}
for the Skyrme term
\begin{eqnarray}
{\cal L}_{\rm sk}&=& -\frac{1}{4 e^2}\left(\mbox{\boldmath $A$}_\mu
\cdot \mbox{\boldmath $A$}^\mu \mbox{\boldmath $A$}_\nu\cdot
 \mbox{\boldmath $A$}^\nu-\mbox{\boldmath $A$}_\mu\cdot 
\mbox{\boldmath $A$}_\nu \mbox{\boldmath $A$}^\mu\cdot 
\mbox{\boldmath $A$}
^\nu\right)
\nonumber \\
&= & 
-\frac{1}{4e^2}
\left\{ 
\frac{\sin^4 F}{F^4}\left[
\left(\partial ^\mu \mbox{\boldmath $F$}\cdot \partial_\mu\mbox{\boldmath $ F$}
\right)\left(\partial^\nu\mbox{\boldmath $F$}
\cdot \partial_\nu \mbox{\boldmath $ F$}
\right)
-\left(\partial ^\mu \mbox{\boldmath $F$}\cdot \partial_\nu \mbox{\boldmath $F$}
\right)\left(\partial^\nu \mbox{\boldmath $ F$}\cdot \partial_\mu 
\mbox{\boldmath $F$}\right)
\right]\right.
\nonumber \\
& & 
+\frac{2\sin^2F(F^2-\sin^2 F)}{F^6} 
\left[
\left(\partial ^\mu \mbox{\boldmath $F$}
\cdot \partial_\mu \mbox{\boldmath $F$}
\right)
\left(\mbox{\boldmath $F$}\cdot\partial_\nu \mbox{\boldmath $F$}\right)
(\mbox{\boldmath $F$}\cdot\partial ^\nu \mbox{\boldmath $F$})
\right.
\nonumber \\
& &
\left.
\left.
-\left(\partial ^\mu\mbox{\boldmath  $F$}
\cdot \partial_\nu \mbox{\boldmath $F$}
\right)
\left(\mbox{\boldmath $F$}\cdot\partial_\mu \mbox{\boldmath $F$}\right)
\left(\mbox{\boldmath $F$}\cdot\partial ^\nu \mbox{\boldmath $F$}\right)
\right]
 \right\},
\end{eqnarray}
and for the pion mass term
\begin{equation}
{\cal L}_{\rm m} = m_\pi^2 f_\pi^2 (\cos F-1).
\end{equation} 

From the results above we can obtain the equation of motion from the
usual starting point,
\begin{equation}
\partial_\mu \frac{\partial \cal L} {\partial (\partial_\mu \mbox{\boldmath 
$F$})}-\frac{\partial \cal L}{\partial \mbox{\boldmath 
$F$}}=0.
\end{equation}
Since the equations of motion are long and complicated, the form 
in terms of  $\mbox{\boldmath $F$}$ are displayed in
the appendix. 

In the next section we discuss the solution of these equations.
Before turning to that section we record two other important formulae.
The baryon number density is given by 
\begin{equation}
{\cal B}=-\frac{1}{2\pi^2}\varepsilon_{\alpha\beta\gamma}
\varepsilon^{i j k} A_i^{\alpha}A_j^{\beta}A_k^\gamma.
\end{equation}
Here $A$ is defined in Eq.\ref{Eq:A};  
 $ijk$ are spatial indices and $\alpha\beta\gamma$ isospin indices.
This density must integrate to zero for a system of Skyrmion and anti-Skyrmion.
The energy density of the system is given by
\begin{equation}
{\cal E}=\frac{f_\pi^2}{2}A_i^\alpha A_i^\alpha+\frac{1}{4e^2}\left[
	(A_i^\alpha A_i^\alpha)^2-(A_i^\alpha A_j^\alpha)(A_i^\beta A_j^\beta)
	\right]+ m_\pi^2 f_\pi^2(1-\cos F). 
\end{equation}
To find the total interaction energy of a Skyrmion and anti-Skyrmion 
configuration, we first solve the equations of motion for the 
static field configuration corresponding to our chosen constraints,
(Skyrmion anti-Skyrmion separation and relative grooming) and then use
that solved field configuration in the integrated energy density
to obtain the interaction energy. 
Note the integrated energy is the total energy of the system. To obtain
the interaction energy or the potentials, one must subtract the 
rest energy of the nucleon and antinucleon from that total energy.
This is done by subtracting two large numbers to obtain what is usually
a much smaller one, and hence requires considerable precision in the 
underlying calculation. Achieving that precision is the goal of the 
full Skyrme calculation and explains why it must be done so
carefully. 

\section{Calculation}
Our aim is to obtain the minimized static energy of the Skyrmion and 
anti-Skyrmion separated by a certain distance and with a definite relative
grooming. 
We need to consider the initial field configuration, the symmetry of 
the field, and  constraints on it before embarking on solving the equations. 
The nonlinear nature of the equation of motion also brings the risk 
of numerical instabilities. These instabilities are more pronounced for 
Skyrmion-antiSkyrmion systems 
than for Skyrmion-Skyrmion due to the lack of a topological constraint
in the Skyrmion-antiSkyrmion case that is due to annihilation.

The essence of our numerical method is to begin with a trial
configuration that has the required symmetries, groomings and
distance constraint. Then to use that configuration as the seed
in the Skyrme equations of motion with a relaxation algorithm.
The system will relax to the lowest energy Skyrme configuration
compatible with the constraints. 
We begin with the symmetrized product ansatz for a groomed Skyrmion and 
a groomed anti-Skyrmion as initial configuration. 
\begin{equation}
U=\frac{1}{N}(U_1 U_2+U_2 U_1),
\end{equation}
where 
\begin{equation}
U_1=A{\cal U}(\vec{r}-\frac{1}{2}R\hat{x})A^{\dagger}
\end{equation}
and
\begin{equation}
U_2=A^\dagger{\cal U}^\dagger(\vec{r}+\frac{1}{2}R\hat{x})A .
\end{equation} 
The SU(2) field ${\cal U}$ is that for a single Skyrmion
in the defensive hedgehog configuration.
The grooming matrix $A=\exp i(\theta/4)\mbox{\boldmath $\tau$}\cdot \hat{a}$
imparts a relative grooming of $\theta$ around direction $\hat{a}$
between the Skyrmion centered at $(R/2)\hat{x}$ and the anti-Skyrmion at
$-(R/2)\hat{x}$.
The normalization factor $N$ makes the symmetrized ansatz unitary.

Three essential groomings are needed to map the interaction energy 
of the Skyrmions into that of baryons. They 
are 1) no grooming; 2) relative grooming of $\pi$ around  the $x$ axis 
(the axis separating the Skyrmion and anti-Skyrmion); and 3)
relative grooming of $\pi$ around the $y$ or 
$z$ axis. Each of these groomings has symmetries that we exploit
in order to simplify our calculation. The reflection symmetries
(in the x, y and z
planes) of the fields are shown in Table 1 for the three situations.
\begin{table}
\begin{center}
\begin{tabular}{|c|c|c|c|}\hline
 &x&y&z \\ \hline\hline
\multicolumn{4}{|c|}{no grooming} \\ \hline
$F_1$ & E &E &E \\ \hline
$F_2$ & O &O &E \\ \hline
$F_3$ & O &E &O \\ \hline \hline
\multicolumn{4}{|c|}{x--$\pi$} \\ \hline
$F_1$ & E &E &E \\ \hline
$F_2$ & E &E &O \\ \hline
$F_3$ & E &O &E \\ \hline \hline
\multicolumn{4}{|c|}{z--$\pi$} \\ \hline
$F_1$ & E &O &E \\ \hline
$F_2$ & O &E &E \\ \hline
$F_3$ & O &E &O \\ \hline
\end{tabular}
\end{center}
\caption{Symmetry of the three groomed configurations. The E and O label
the {\em even} or {\em odd} symmetry of the particular field component across
the x, y or z plane.}
\end{table}

These three grooming are the normal modes for the Skyrmion and 
anti-Skyrmion system and their symmetries should be maintained 
in solving the equation of motion. We achieve this by restricting our
calculation to the first octant of the full
three dimensional space, and use the reflection symmetries 
as boundary conditions for the field at the x, y and z planes.
This restriction also reduces our computational time by a factor
of eight. 

The anti-Skyrmion and Skyrmion system is not stable under the equation of
motion unless it is subjected to a proper constraint. Since the total baryon 
number is zero for this system, the usual topological constraint
used in the multi-Skyrmion case is not present here. For the Skyrmion-Skyrmion, 
the distance $R$ between the two, as defined in \cite{W&W} by
\begin{equation} 
\frac{1}{4}R^2=\int d^3r {\cal B}(\vec{r})r^2
\end{equation}
is constrained to a definite value. For the Skyrmion-antiSkyrmion, 
this can no longer be used. The baryon density is odd across
the $x$-plane and hence the
distance $R$ obtained from such
a density is obviously zero.  The constraint we propose instead 
is a topological one. When the separation (as defined in the
symmetrized product ansatz) is more than $0.8$ fm, the 
absolute value of the chiral angle always has a peak value of $\pi$ at
two places on the x-axis for all three of the 
groomings. Note that these locations are not 
$\pm(R/2)\hat{x}$, although they are asymptotically for 
very large separation. At these two locations, we hold the field values 
constant as defined in the symmetrized product ansatz 
when we solve the equation of motion. We use the separation value of
$R$ as defined in the symmetrized product ansatz.  For distance less
than $0.8$ fm, the solution becomes numerically less tractable and a definition
of separation is unclear. We do not study interaction energy of these cases 
in the present paper.
Such small separations are not meaningful for the nucleon antinucleon
problem since empirically one finds complete annihilation well before
that distance is reached. As we shall see below, we also find very
strong coupling to annihilation below 1 fm. 

Using the symmetry and topological constraints discussed above, we 
solve the equations of motion in the first octant of coordinates to 
obtain the minimized energy. To do so, we reduce the equation of motion to
a dissipative one by setting all first order time derivatives to
zero at each time step. The resulting Langevin type equation is 
solved on a lattice in 
the first octant with appropriate boundary conditions on the faces
that form the boundaries of the octant. We propagate 
the field in discrete time steps until 
a stable final field configuration is reached. The field 
profile is strongly peaked around the locations where the chiral angle
reaches the value of $\pi$. We use a lattice coordinate 
system with variable
grid that has the densest 
distribution of points at these locations.

\section{Results}
In this section we present the  
results of our numerical calculations of
the Skyrmion-antiSkyrmion interaction and of the corresponding
nucleon-antinucleon interaction. We begin with the interaction energy of
the Skyrmion-antiSkyrmion ($S \bar{S}$) system in the three groomings 
required to extract the nucleonic potentials, namely the $S \bar{S}$ 
system with no relative grooming, the system with a relative grooming
of $\pi$ around the axis joining the $S$ and $\bar{S}$ (the x-axis)
and the system with a relative grooming of $\pi$ along an axis at
right angles to the line connecting the $S$ and $\bar{S}$ (z-axis).
The results of our calculation are shown in Figure 1. The Figure
shows the total energy of the $S \bar{S}$ system as a function of
separation distance. Note that for large separation the total energy
is that of two free Skyrmions, namely 2.92 GeV for our choice of
parameters. The separation distance is
defined as discussed in Section 3. It includes a scheme for 
holding the pion field at its maximum value at particular points. 
We only show a portion of the
separation distance in the figure since for distances larger than
2.2 fm, the interaction is nearly zero and for distances below 0.8 fm,
our entire scheme as well as the meaning of adiabatic interaction 
ceases to make sense. It is important to note that the ``holding" process
is essential to our calculation, since without it all $S$ $\bar{S}$ 
configurations would relax to the lowest total energy of such a system,
namely zero corresponding to total annihilation. In this sense our
calculation is very different from that of Walhout and Wambach for
the $S-S$ system \cite{W&W}.

From Figure 1 we see that the x-grooming
channel (grooming along the axis joining the two) is attractive while
grooming along the orthogonal axis is repulsive. This is just the opposite
of what happens for the $S-S$ system, as we might expect from g-parity
or related arguments. The no grooming channel is also 
repulsive. Well below 0.8 fm it must turn 
attractive again since for zero separation and no grooming  
there is complete annihilation and the total energy must be zero.
The results in Figure 1 should be compared with our previous
results calculated with the product ansatz \cite{LuRDA96}. The 
qualitative results are the same. In particular the results
agree at large distances where we expect the product ansatz to
be a good approximation. However the full calculation with 
dynamical relaxation of the fields has much more attraction in the
attractive channel and less repulsion in the repulsive channel
as compared with the product ansatz. All this
suggests, correctly as we will see below, that the full calculation will
yield the strong $N-\bar{N}$ midrange 
attraction seen phenomenologically, but
missing in the product ansatz.

The full dynamical results shown in Figure 1 suggest
that for the $S \bar{S}$ system with no grooming there is a
point of 
unstable equilibrium inside 0.8 fm. If the $S \bar{S}$ system is released
from rest just inside that unstable point, it will proceed to annihilate, while
if it is released just outside, it will separate to infinity. Such a critical
radius should continue to exist at scattering energies such that at 
impact parameters less than a certain value annihilation occurs while for
impact parameters greater than that value, the $S$ and $\bar{S}$ emerge
at infinity. It should be emphasized that this is a classical scattering
problem, so that no probabilities are allowed and that, due to the topology,
the Skyrmion must appear at infinity in its entirety or not at all. The 
existence of such a critical impact parameter seems not to have been noted
before, and reveals the existence of a new singular length in the Skyrme
model. Scattering exactly at the 
singular impact parameter must involve
very long time delays. 

Some sense of the significance of the three groomings can be gathered
by looking at arrow plots. These are plots of the chiral angle field,
$\mbox{\boldmath $F$}$, at various points in a plane through the 
$S$ and $\bar{S}$. 
We use the x-y plane. Recall that $F \rightarrow 0$ at large distances
and at all places where the energy density is small. We show the arrow plots
for the three  groomings, always with a $S \bar{S}$ separation of
1 fm. in Figure 2. Figure 2a shows the no grooming case. The $S$ hedgehog
is on the left and the $\bar{S}$ anti-hedgehog on the right. The 
attractive channel, with grooming along the x-axis that joins the $S$ and
$\bar{S}$ is shown in Figure 2b. It is clear that most of the 
$\mbox{\boldmath $F$}$
arrows are now quite short, corresponding to reduced energy  density and
thus attraction. Finally the repulsive channel with grooming along the 
z-axis is shown in Figure 2c. Here not only do the arrows not get short,
but it is clear that as the $S$ and $\bar{S}$ approach the opposite 
arrows will clash leading to repulsive energy at short distances.

We now turn to extracting the nucleon antinucleon interaction from the
$S \bar{S}$ results. As we have emphasized before, \cite{LuRDA96,WA},
there are two steps in this process. First from combinations of the
various $S \bar{S}$ groomings we can construct matrix elements of the
baryon anti-baryon interaction. These include not just the nucleon-antinucleon
matrix elements but also matrix elements coupling to the $\Delta$ resonance
and diagonal matrix elements involving the $\Delta$'s. The observed 
nucleon-antinucleon interaction involves these coupled terms since all
that we are certain of experimentally is that there are nucleons and
anti-nucleons asymptotically. When they get close and begin to interact,
$\Delta$'s are allowed to mix in. The formalism for including that mixing using
the Born-Oppenheimer approximation has been presented in detail before,
\cite{LuRDA96,WA}, and we do not repeat it here. 
Rather we just 
give the results for the various nucleon antinucleon interactions showing
both the results from taking only the nucleonic projections and the 
results from including the full mixing. We will see, as we expect,  that
the diagonalized Born-Oppenheimer mixing 
leads to enhanced attraction. We also include some finite $N_C$ corrections
for $N_C=3$ in our calculation, as we did before. 

Figure 3 shows the nucleon antinucleon central interaction in the $T=0$
channel. We show the result from taking only the nucleon projection as well
as the result from the full Born-Oppenheimer diagonalization. 
Both show much more attraction than we found in the case of the
product ansatz, with the diagonalization significantly 
increasing the attraction. With diagonalization, 
the potential is about as strong as that seen phenomenologically either
in the Bryan-Phillips analysis, \cite{BP}, or in that of the Nijmegen 
group, \cite{Nijmegen}, but the dependence on $R$ is
somewhat different. In all cases, the potential is very strong. 
 Note that 
the potentials do not mean much inside of 1 fm since there is very strong
absorption there due to annihilation. Figure 4 shows the same results for the
central but $T=1$ channel. Due to the much greater effect of $\Delta$ mixing in
this channel, the effect of diagonalization is to produce much more
attraction. Now the full calculated potential is comparable to 
or perhaps even more attractive than the 
phenomenological results, but most of the overshoot is at the
short distances where the potential concept is suspect. 
Again we find the correct scale (about half of the scale of the
$T=0$ case) of the central midrange attraction 
and again far more attraction than we
found in the product ansatz case. The spin dependent potentials are shown
in Figures 5 and 6. For the $T=0$ case the agreement with the 
phenomenological results is satisfactory. The product ansatz calculation
gave the completely wrong answer for this case. For the $T=1$ spin
dependent potential, the results are not as good,  but the 
smallness of the potential is reproduced.
In our calculation, that small value arises from the cancelation of 
a number of large factors and hence is very sensitive to details.    
The $T=0$ tensor force (Figure 7)
is in nearly perfect agreement with the phenomenology  
and the $T=1$ tensor force (Figure 8) is rather well given,
particularly at the larger distances. The tensor force
is dominated by one pion exchange and hence was also well
accounted for in the product ansatz.

Thus far we have concentrated on the nucleon antinucleon interaction
potentials. But a defining feature of the baryon anti-baryon system
is annihilation. Phenomenologically this can be represented in 
the initial state by an optical or absorptive potential. The
data suggests that that potential is very strong, but short
ranged. That is it sets in sharply at around 1 fm\cite{AmMy}. In the 
classical Skyrme picture we would expect to see this as a sudden
drop in the local baryon number. That is, consider the baryon number
in the half space where the Skyrmion sits. Asymptotically that
number should be one. Due to the finite size of our lattice, we would
find a number slightly less than one. In Figure 9 we plot that 
baryon number as a function of Skyrmion anti-Skyrmion separation
for the three groomings. We see that it is indeed very near one
for large separations. In the repulsive channel it remains near one
even for relatively small separations (1 fm). In the attractive
channel we see something very different. The baryon number begins
near one at large separation, but near 1.2 fm it plunges abruptly to
a value very near zero. This is annihilation. The fact that it occurs
so sharply and at roughly the distance required by the data give
further support for the Skyrme picture. In the channel with no grooming
there is an even sharper onset of annihilation, but at a somewhat
smaller distance. In our subsequent work we will use this 
annihilation mechanism combined with a coupled channel quantum approach
using the interaction potentials we have obtained here to model
the initial state.

All in all the nucleon antinucleon interaction extracted from the full,
dynamical Skyrmion anti-Skyrmion interaction agrees well with
the major trends of the
data, and in particular reproduces the strong central attraction seen in the
region between 1 and 2 fm, and the rapid onset of annihilation. 
This demonstrates that the Skyrme approach can account for the nucleon
antinucleon initial state, as well as the final annihilation state. 

\section{Discussion}

We have shown that the strong central attraction, 
the sharp onset of annihilation at about 1 fm and the 
other the principal features of the phenomenological nucleon antinucleon
interaction emerge from a careful calculation of that interaction based
on the Skyrme model. Since the Skyrme picture models QCD in the classical
or large number of colors and  low energy 
limit and since that limit is the  appropriate one
for low energy nonperturbative phenomena, our calculation links low
energy nucleon antinucleon interactions to QCD. There are two essential
pieces to our calculation. The first is a careful and thorough
dynamical (though static) computation of the Skyrmion anti-Skyrmion
configurations. It is this step that is vital to getting the strong,
midrange attraction missing from the product ansatz approach to the same
problem. This dynamical calculation is complex and difficult and 
represents the major new work presented here. 
The second step involves using not just Skyrmion to nucleon
projections but also state mixing to obtain the full nucleon interaction.
We have used these two parts to obtain a very satisfactory account of
the nucleon anti-nucleon  interaction.

Previously we have shown that the Skyrme model can account for the major
features of the annihilation branches in low energy nucleon antinucleon
annihilation, \cite{batch2}, but  that calculation had no initial state
dynamics. In this paper we have shown that the initial state can also
be successfully studied using the Skyrme approach. Our next step is
to combine these two and give a complete, QCD based (via Skyrme) description
of low energy nucleon antinucleon annihilation from start to finish.
We plan to turn to that task next.

{\noindent \bf Acknowledgement}

The authors wish to thank Dr.~Niels R.~Walet for his interest
in this project and in particular for his help in emphasizing
the importance of symmetries in simplifying the calculation.
This work was supported in part by a grant from the National
Science Foundation.

\appendix
\section{Equation of motion}
The equation of motion in the chiral angle $\mbox{\boldmath  $F$}$ is
a long expression. It contain three terms.
The term from the nonlinear sigma model is
\begin{eqnarray}
f_\pi^2 &&\left[ 
\partial_\mu\partial^\mu \mbox{\boldmath 
$F$} \frac{\sin^2F}{F^2} \right. 
\nonumber \\
&+& \mbox{\boldmath $F$} \left(
\mbox{\boldmath $F$}\cdot\partial_\mu\partial^\mu \mbox{\boldmath 
$F$}\right) \frac{F^2-\sin^2 F}{F^4}
\nonumber \\
&+& \mbox{\boldmath $F$} \left(\partial^\mu\mbox{\boldmath $F$}
\cdot\partial_\mu\mbox{\boldmath $F$}\right)
\frac{F-\sin F \cos F}{F^3}
\nonumber \\
&+& 2 \partial_\mu \mbox{\boldmath $F$} \left(
\mbox{\boldmath $F$}\cdot\partial^\mu \mbox{\boldmath $F$}\right)
\frac{\sin F( F\cos F -\sin F)}{F^4}
\nonumber \\
&+& \mbox{\boldmath $F$} \left. \left(
\mbox{\boldmath $F$}\cdot\partial^\mu \mbox{\boldmath $F$}\right)
\left(\mbox{\boldmath $F$}\cdot\partial_\mu 
\mbox{\boldmath $F$}\right)
\frac{2\sin ^2 F -F^2-F \sin F \cos F}{F^6}
\right].
\end{eqnarray}

The term from the Skyrme term is
\begin{eqnarray}
-\frac{1}{e^2}\left\{ \right.
& & \frac{\sin^4 F}{F^4} \left[
\partial_\mu\partial^\mu \mbox{\boldmath $F$}
(\partial^\nu \mbox{\boldmath $F$}\cdot
\partial_\nu \mbox{\boldmath $F$})
-\partial_\mu\partial^\nu \mbox{\boldmath $F$}
(\partial^\mu \mbox{\boldmath $F$}\cdot
\partial_\nu \mbox{\boldmath $F$})\right]
\nonumber \\
&+ & \frac{\sin^2 F(F^2 - \sin^2 F)}{F^6}
\left[\partial_\mu\partial^\mu \mbox{\boldmath $F$}
(\mbox{\boldmath $F$}\cdot \partial_\nu \mbox{\boldmath $F$})
(\mbox{\boldmath $F$}\cdot \partial^\nu \mbox{\boldmath $F$})
-\partial_\mu\partial^\nu \mbox{\boldmath $F$}
(\mbox{\boldmath $F$}\cdot \partial^\mu \mbox{\boldmath $F$})
(\mbox{\boldmath $F$}\cdot \partial_\nu \mbox{\boldmath $F$})
\right. \nonumber \\
&+& \partial^\mu \mbox{\boldmath $F$}\left(
(\mbox{\boldmath $F$}\cdot\partial_\mu\partial_\nu \mbox{\boldmath $F$}
\partial^\nu \mbox{\boldmath $F$}-(\mbox{\boldmath $F$}\cdot
\partial^\nu\partial_\nu \mbox{\boldmath $F$})
((\mbox{\boldmath $F$}\cdot \partial_\mu \mbox{\boldmath $F$})
\right) \nonumber \\
&+& \mbox{\boldmath $F$}\left(
(\mbox{\boldmath $F$}\cdot \partial^\mu \partial_\mu \mbox{\boldmath $F$})
(\partial^\nu\mbox{\boldmath $F$}\cdot\partial_\nu \mbox{\boldmath $F$})
-(\mbox{\boldmath $F$}\cdot \partial^\mu\partial_\nu \mbox{\boldmath $F$})
(\mbox{\boldmath $F$}\cdot \partial_\mu\partial^\nu \mbox{\boldmath $F$
})       \right) \nonumber \\
&+& \mbox{\boldmath $F$}\left(
(\mbox{\boldmath $F$}\cdot \partial^\mu \mbox{\boldmath $F$
})
(\partial^\nu\mbox{\boldmath $F$}\cdot\partial_\mu\partial_\nu \mbox{\boldmath $F$})
-(\mbox{\boldmath $F$}\cdot \partial^\mu\mbox{\boldmath $F$})
(\partial_\mu\mbox{\boldmath $F$}\cdot \partial_\nu\partial^\nu \mbox{\boldmath $F$})
\right)\left.\right]
\nonumber \\
&+& \left[\frac{\sin^2 F}{F^4}-\frac{\sin^3 F\cos F}{F^5}\right]
\mbox{\boldmath $F$}\left[
\left(\partial^\mu\mbox{\boldmath $F$}\cdot\partial_\mu\mbox{\boldmath $F$}\right)
\left(\partial^\nu\mbox{\boldmath $F$}\cdot\partial_\nu\mbox{\boldmath $F$}\right)-\left(\partial^\mu\mbox{\boldmath $F$}\cdot\partial_\nu\mbox{\boldmath $F$}\right)
\left(\partial^\nu\mbox{\boldmath $F$}\cdot\partial_\mu\mbox{\boldmath $F$}\right)\right]
\nonumber \\
&+&\left[4\frac{\sin^3 F \cos F}{F^5}-3\frac{\sin^4 F}{F^6}
-\frac{\sin^2 F}{F^4}\right] \nonumber \\
&&
\partial^\mu\mbox{\boldmath $F$}\left[
\left(\mbox{\boldmath $F$}\cdot\partial_\mu\mbox{\boldmath $F$}\right)
\left(\partial_\nu\mbox{\boldmath $F$}\cdot\partial^\nu\mbox{\boldmath $F$}
\right)
-
\left(\mbox{\boldmath $F$}\cdot\partial_\nu\mbox{\boldmath $F$}\right)
\left(\partial_\mu\mbox{\boldmath $F$}\cdot\partial^\nu\mbox{\boldmath $F$}
\right)\right]
\nonumber \\
&+&
\left[ \frac{\sin F \cos F}{F^5}-2\frac{\sin^3 F \cos F}{F^7} - 2\frac{\sin^2 
F}{F^6}+3 \frac{\sin^4 F}{F^8}\right]
\nonumber \\
&&
\mbox{\boldmath $F$}
\left[
\left(\partial_\mu\mbox{\boldmath $F$}\cdot\partial^\mu\mbox{\boldmath $F$}
\right)
\left(\mbox{\boldmath $F$}\cdot\partial_\nu\mbox{\boldmath $F$}\right)
\left(\mbox{\boldmath $F$}\cdot\partial^\nu\mbox{\boldmath $F$}\right)
-
\left(\partial_\mu\mbox{\boldmath $F$}\cdot\partial^\nu\mbox{\boldmath $F$}
\right)
\left(\mbox{\boldmath $F$}\cdot\partial_\nu\mbox{\boldmath $F$}\right)
\left(\mbox{\boldmath $F$}\cdot\partial^\mu\mbox{\boldmath $F$}\right)\right]
\left. \right\},
\end{eqnarray}
and finally the term with the pion mass is
\begin{equation}
- m_\pi^2 f_\pi^2\frac{\sin F}{F}\mbox{\boldmath $F$}.
\end{equation}
The equation of motion is
given by equating the sum of these three terms to zero. 

\begin{figure}
\centerline{\hbox{
\psfig{figure=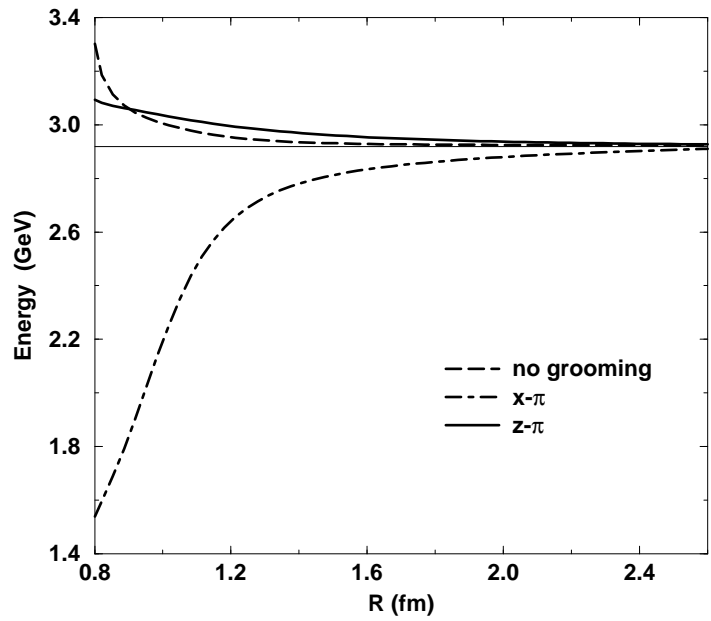,height=3.7in}
}}

\caption{Total energy of the $S \bar{S}$ system as a function of
separation distance for three major groomings. The horizontal
line represents twice the Skyrmion mass, at 2.92 GeV.
}\label{Fig.1}
\end{figure}
\newpage

\begin{figure}
\centerline{\hbox{
\psfig{figure=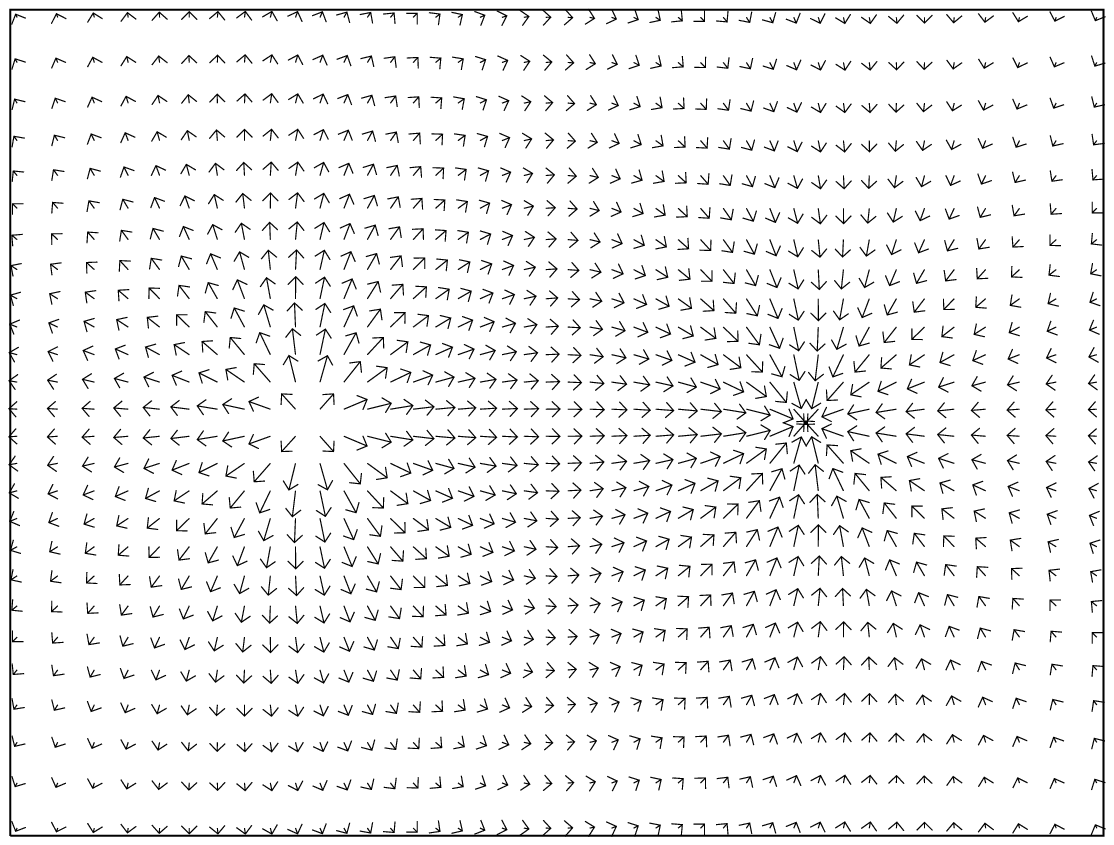,height=2.2in}
}}
\vspace*{-0.6cm}
\begin{center}
2a 
\end{center}

\centerline{\hbox{
\psfig{figure=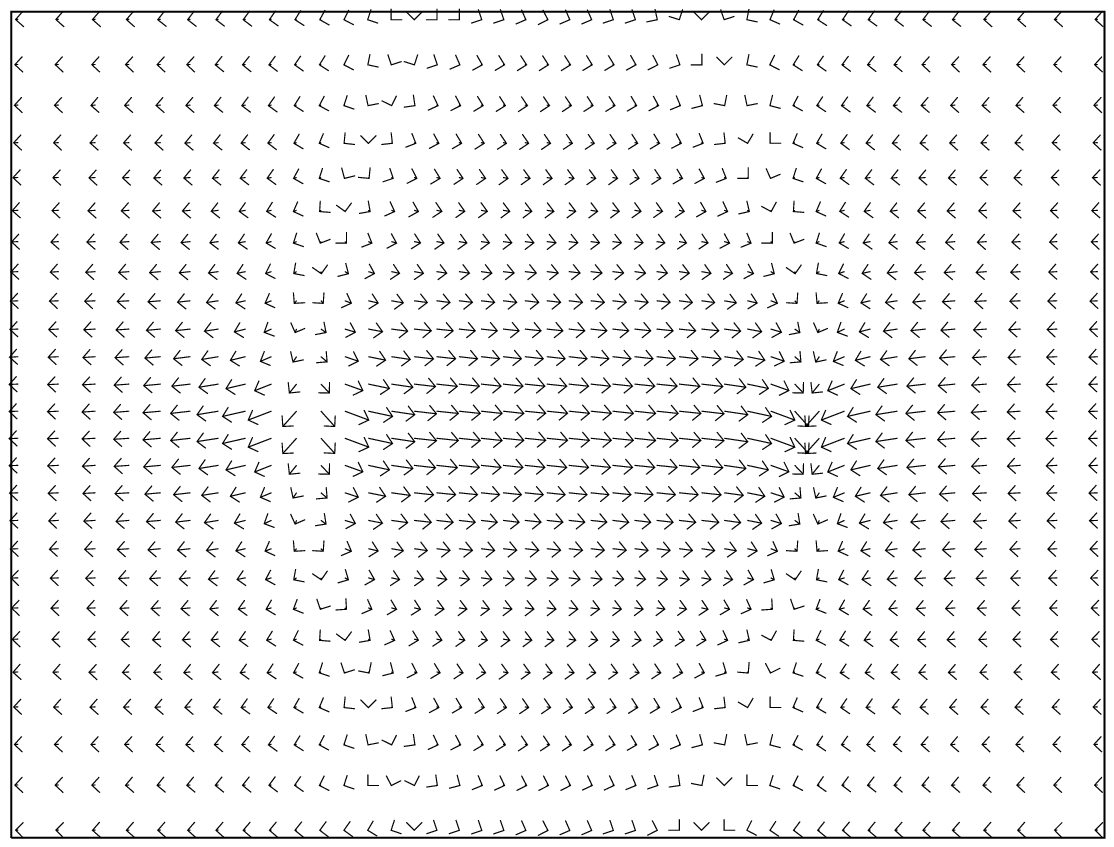,height=2.2in}
}} 
\vspace*{-0.6cm}
\begin{center}
2b 
\end{center}

\centerline{\hbox{
\psfig{figure=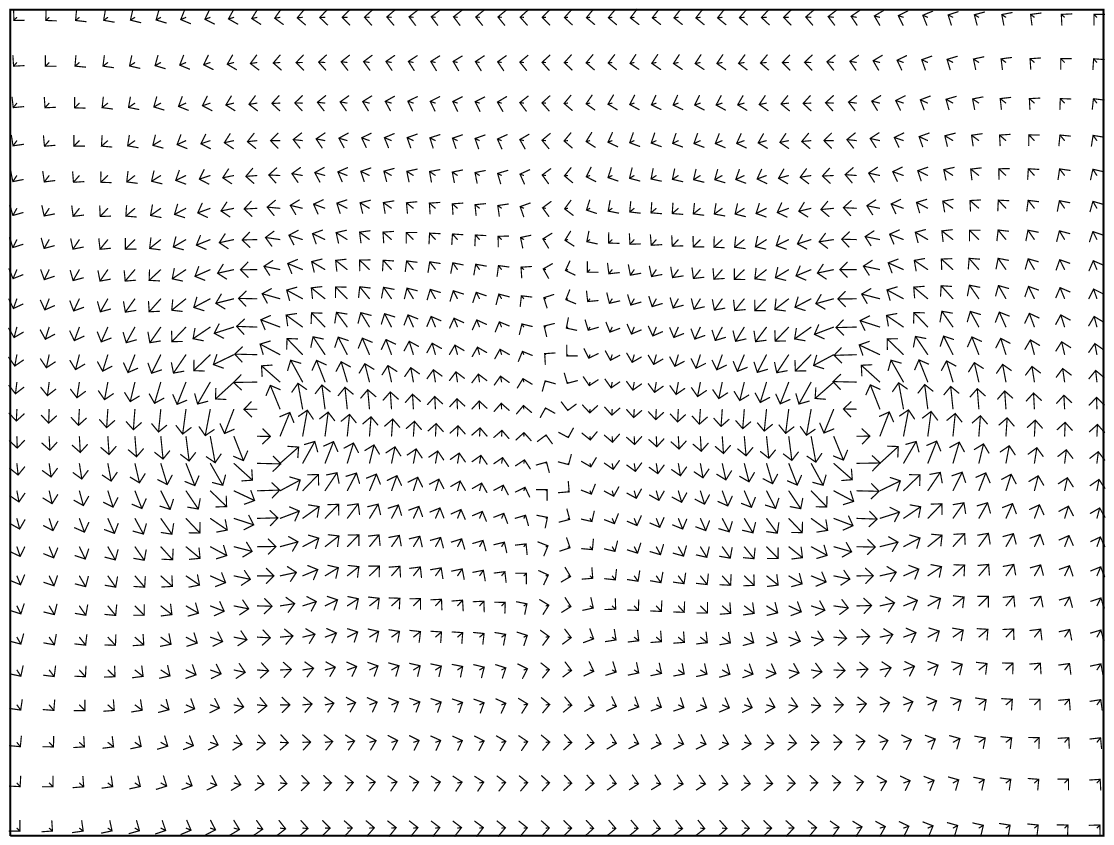,height=2.2in}
}} 
\vspace*{-0.6cm}
\begin{center}
2c 
\end{center}

\caption{Arrow plots of the chiral angle field for the three groomings at 
separation of 1 fm.
Figure 2a shows the no grooming case, 2b for grooming of $\pi$ around
x-axis and 2c grooming of $\pi$ around z-axis.
The Skyrmion is on the left and the anti-Skyrmion on the right.}
\label{Fig.2}
\end{figure}


\begin{figure}
\centerline{\hbox{
\psfig{figure=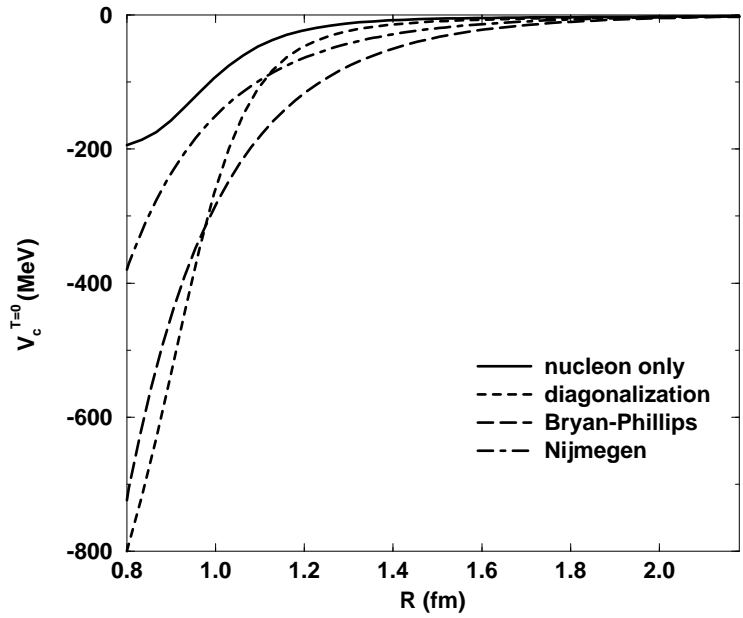,height=3.7in}
}}

\caption{Central potential $V_C^T$ as a function of $R$ in the
region 0.8--2.2 fm for  the
$T=0$ channels. The solid line gives the nucleons only result from
minimization. The short dashed line is the result of the state
mixing using
full Born-Oppenheimer diagonalization.  The phenomenological potentials 
based on meson exchange 
are shown by the long-dashed line for Bryan-Phillips potential
{\protect{\cite{BP}}} and by the dash-dotted line for the
Nijmegen potential {\protect{\cite{Nijmegen}}}.
}
\label{Fig.3}
\end{figure}


\begin{figure}
\centerline{\hbox{
\psfig{figure=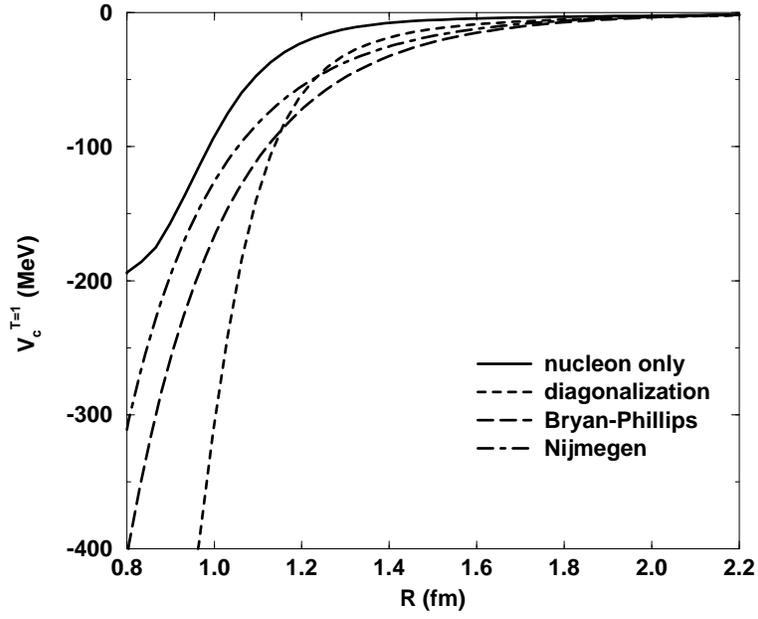,height=3.7in}
}}
\caption{Central potential, same as in Fig.~3 but for $T=1$.}
\label{Fig.4}
\end{figure}


\begin{figure}
\centerline{\hbox{
\psfig{figure=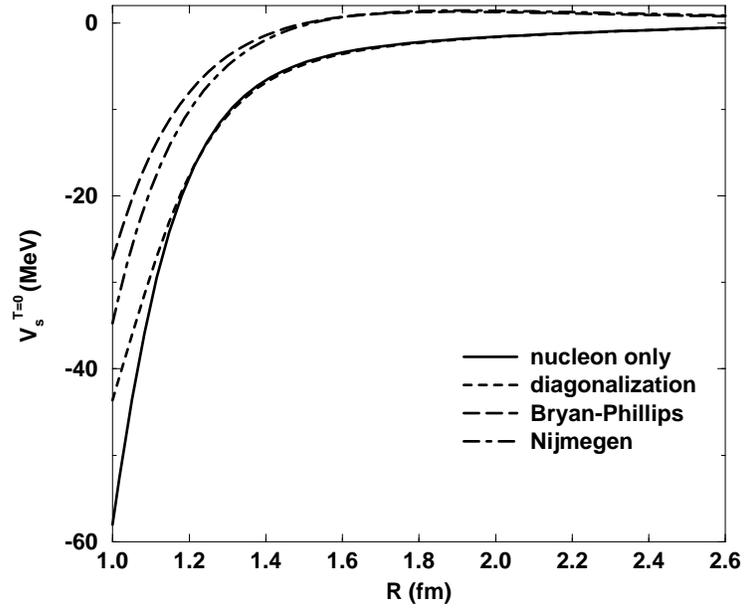,height=3.7in}
}}

\caption{The spin dependent potential $V_s$ as a function of $R$
in the region 1--2.6 fm for
$T=0$. Labeling of curves is the same as in Fig.~3.}
\label{Fig.5}
\end{figure}


\begin{figure}
\centerline{\hbox{
\psfig{figure=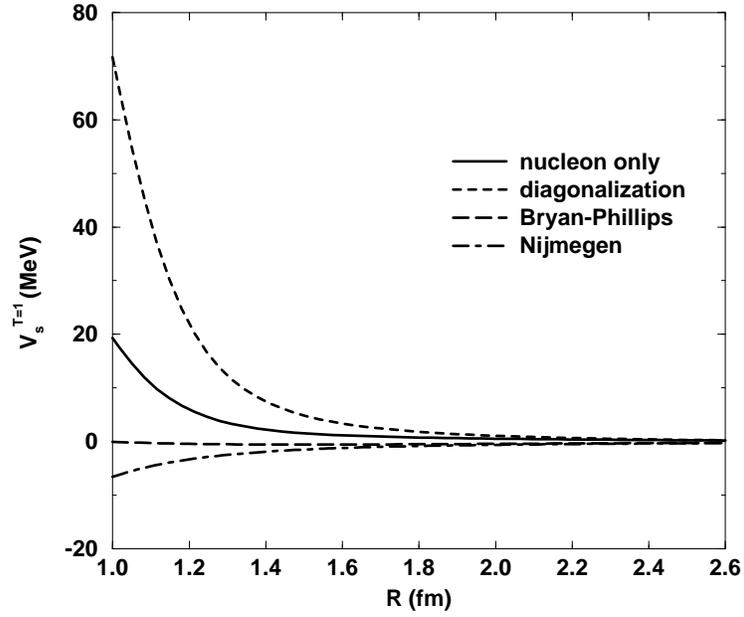,height=3.7in}
}}
\caption{Spin-dependent potential, same as Fig.~5
but for $T=1$.}
\label{Fig.6}
\end{figure}
\begin{figure}
\centerline{\hbox{
\psfig{figure=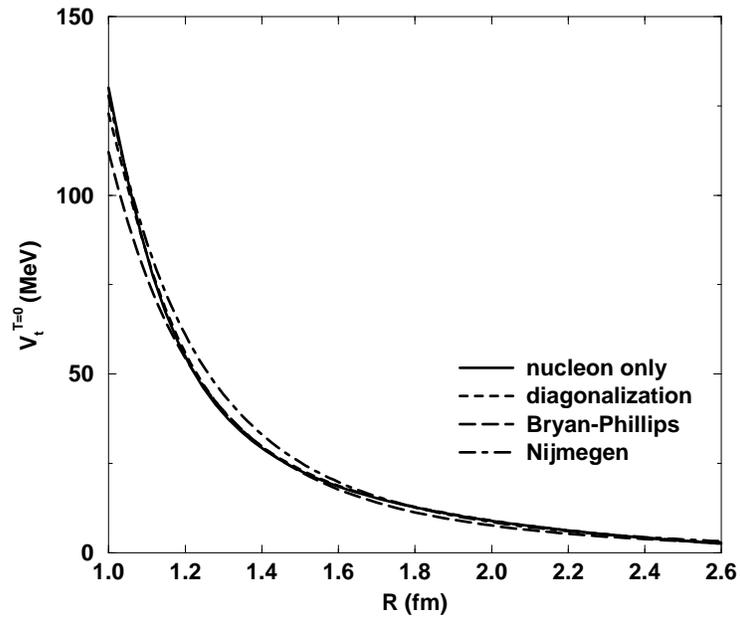,height=3.7in}
}}

\caption{Tensor potential $V_t$ as a function of $R$
in the region 1--2.6 fm for $T=0$.
Labeling of curves is the same as in Fig.~3.
}\label{Fig.7}
\end{figure}

\begin{figure}
\centerline{\hbox{
\psfig{figure=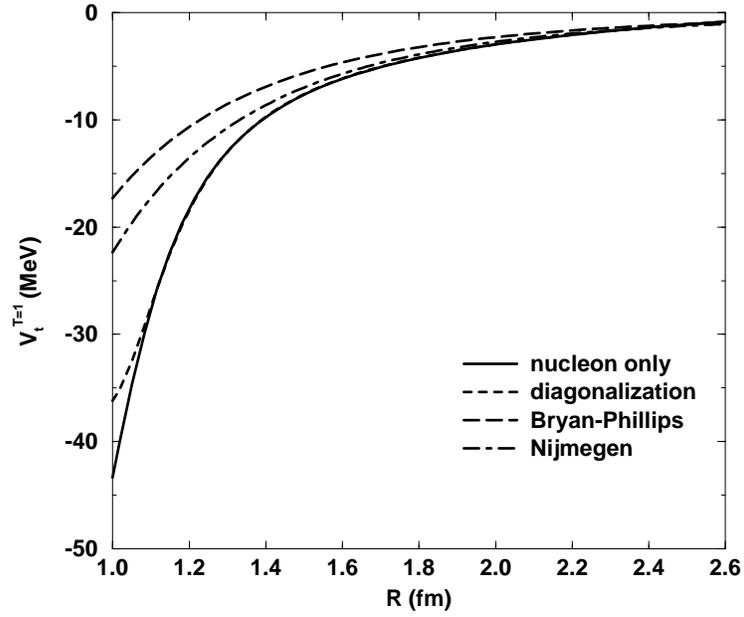,height=3.7in}
}}

\caption{Tensor potential, same as in Fig.~7 but for $T=1$.
 }\label{Fig.8}
\end{figure}

\begin{figure}
\centerline{\hbox{
\psfig{figure=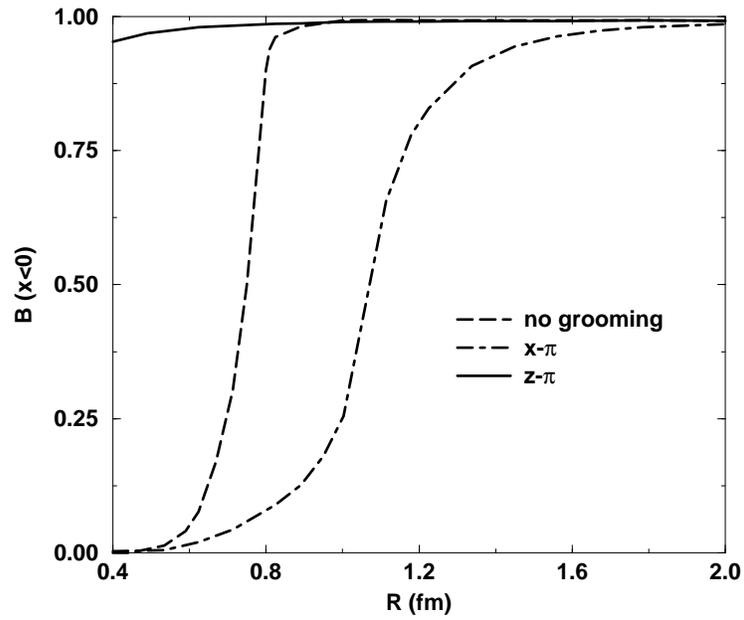,height=3.7in}
}}

\caption{Baryon number in the half of space (x$<$0) where 
the Skyrmion resides, for three major groomings, as a function
of separation.
 }\label{Fig.9}
\end{figure}
\end{document}